\documentclass{article}%
\usepackage{amsmath}
\usepackage{amsfonts}
\usepackage{amssymb}
\usepackage{graphicx}%
\providecommand{\U}[1]{\protect\rule{.1in}{.1in}}
\newtheorem{theorem}{Theorem}
\newtheorem{acknowledgement}[theorem]{Acknowledgement}

\newtheorem{remark}[theorem]{Remark}

\begin{document}

\title{ Bosonization of Fermionic Fields and Fermionization of Bosonic Fields}
\author{Waldyr A. Rodrigues Jr.\\Institute of Mathematics, Statistics and Scientific Computation\\IMECC-UNICAMP\\walrod@ime.unicamp.br}
\date{November 15 2016}
\maketitle

\begin{abstract}
In this paper using the Clifford and spin-Clifford bundles formalism we show
how Weyl and Dirac equations formulated in the spin-Clifford bundle may be
written in an equivalent form as generalized Maxwell like form formulated in
the Clifford bundle. Moreover, we show how Maxwell equation formulated in the
Clifford bundle formalism can be written as an equivalent equation for a
spinor field in the spin-Cillford bundle. Investigating the details of such
equivalences this exercise shows explicitly that a fermionic field is
equivalent (in a precise sense) to an equivalence class of well defined boson
fields and that a bosonic field is equivalent to a well defined equivalence
class of fermionic fields These equivalences may be called the bosonization of
fermionic fields and the fermionization of bosonic fields.

\end{abstract}

\section{Introduction}

The idea to bosonize fermion fields and fermionize boson fields is probably a
very old one and has produced in our opinion a considerable amount of
misunderstandings. Here, we show how using the Clifford and spin-Clifford
bundle formalisms we can easily show how to associate to a spin $1/2$
fermionic field an equivalence class of bosonic fields and also how to
associate to a bosonic field an equivalence class of fermionic fields. In
particular the last exercise shows in a trivial way the origin of the many and
many proposed Dirac like representations of the Maxwell field that appeared in
the literature. And before you continue the reading and advertisement is in
order: the mathematical equivalences proved in this paper, of course, do not
change the statistics of sets of particles described by the different classes
(fermions and bosons) of fields.

\section{The Necessary Mathematical Tools}

\subsection{Forget Matrices}

In order to prove in a easy way the results mentioned in the title of this
paper the first thing to be done is to forget the matrix representation of
spinor fields. So, we start recalling the necessary tools.

\textbf{1.} In Minkowski spacetime\footnote{Minkowski spacetime is the
structure $(M,\boldsymbol{\eta},D,\tau_{\boldsymbol{\eta}},\uparrow$), where
$M\simeq\mathbb{R}^{4}$. Moreover $\boldsymbol{\eta}$ is a metric field of
signature $(1,3)$, $D$ is the Levi-Civita connection of $\boldsymbol{\eta}$
and such that its associate Riemann curvature tensor is null. The structure is
oriented by $\tau_{\boldsymbol{\eta}}\in\sec%
{\textstyle\bigwedge\nolimits^{4}}
T^{\ast}M$ and time oriented by $\uparrow$. Details in \cite{rc2007}.} with
manifold $M\simeq\mathbb{R}^{4}$ introduce global coordinates $\{\mathrm{x}%
^{\mu}\}$. Write $\{\mathrm{\partial}_{\mu}=\frac{\partial}{\partial
\mathrm{x}^{\mu}}\}$ and $\{\gamma^{\mu}:=d\mathrm{x}^{\mu}\}$ for the basis
of $TM$ (tangent bundle) and $T^{\ast}M\simeq%
{\textstyle\bigwedge\nolimits^{1}}
T^{\ast}M$ (cotangent bundle)

\textbf{2. }The metrics of the tangent and cotangent bundles are:%
\begin{equation}
\boldsymbol{\eta}=\eta_{\mu\nu}\gamma^{\mu}\otimes\gamma^{\nu}%
,~~~~\mathtt{\eta}=\eta^{\mu\nu}\partial_{\mu}\otimes\partial_{\nu} \label{1}%
\end{equation}

\textbf{3. }The bundle of (non homogeneous) differential form fields is
denoted $%
{\textstyle\bigwedge}
T^{\ast}M$ and%
\begin{equation}%
{\textstyle\bigwedge}
T^{\ast}M=%
{\textstyle\sum\nolimits_{r=0}^{4}}
\oplus%
{\textstyle\bigwedge\nolimits^{r}}
T^{\ast}M, \label{3}%
\end{equation}
where $%
{\textstyle\bigwedge\nolimits^{r}}
T^{\ast}M$ is the bundle of $r$-form fields.

\textbf{4. }For each $x\in M$ the $2^{4}$-dimensional vector space $%
{\textstyle\bigwedge}
T_{x}^{\ast}M$ is isomorphic to the basic vector space of the Clifford bundle
of differential forms $\mathcal{C\ell}(M,\mathtt{\eta})$ for each $x\in M$.
This means that
\begin{equation}%
{\textstyle\bigwedge}
T^{\ast}M\hookrightarrow\mathcal{C\ell}(M,\mathtt{\eta}) \label{4}%
\end{equation}

\textbf{5. }Then, if we suppose that the $\gamma^{\mu}\in\sec%
{\textstyle\bigwedge\nolimits^{1}}
T^{\ast}M\hookrightarrow\sec\mathcal{C\ell}(M,\mathtt{\eta})$ these objects
satisfy the basic relation (where the Clifford product is simply denoted by
juxtaposition)%
\begin{equation}
\gamma^{\mu}\gamma^{\nu}+\gamma^{\nu}\gamma^{\mu}=2\eta^{\mu\nu} \label{5}%
\end{equation}

\textbf{6.} In arbitrary coordinates $\{x^{\mu}\}$ for $U\subset M$ the
metrics of the tangent and cotangent bundles read:%
\begin{equation}
\boldsymbol{\eta}=g_{\mu\nu}dx^{\mu}\otimes dx^{\nu},~~~~\mathtt{\eta}%
=g^{\mu\nu}\frac{\partial}{\partial x^{\mu}}\otimes\frac{\partial}{\partial
x^{\nu}} \label{6}%
\end{equation}
and for the Levi-Civita connection \ $D$ of $\boldsymbol{\eta}$ we have
\begin{equation}
D_{\frac{\partial}{\partial x^{\mu}}}\frac{\partial}{\partial x^{\nu}}%
=\Gamma_{\cdot\mu\nu}^{\alpha\cdot\cdot}\frac{\partial}{\partial x^{\alpha}%
},~~~~D_{\frac{\partial}{\partial x^{\mu}}}dx^{\alpha}=-\Gamma_{\cdot\mu\nu
}^{\alpha\cdot\cdot}dx^{\nu}, \label{7}%
\end{equation}
where in general the Christofell symbols $\Gamma_{\cdot\mu\nu}^{\alpha
\cdot\cdot}$ are not null. However take into account that in the coordinates
$\{\mathrm{x}^{\mu}\}$ it is%
\begin{equation}
D_{\mathrm{\partial}_{\mu}}\partial_{\nu}=0,~~~~D_{\mathrm{\partial}_{\mu}%
}d\mathrm{x}^{\alpha}=0. \label{7a}%
\end{equation}

\textbf{7. }It is very important to take in mind that even if we use general
coordinates $\{x^{\mu}\}$ for $U\subset M$ it is always possible to introduce
in $U$ a set of orthonormal basis for $TM$ and $T^{\ast}M\simeq%
{\textstyle\bigwedge\nolimits^{1}}
T^{\ast}M$. Indeed, defining%
\begin{equation}
e_{\mathbf{a}}=h_{\mathbf{a}}^{\mu}\frac{\partial}{\partial x^{\mu}%
},~~~\mathfrak{g}^{\mathbf{a}}=h_{\nu}^{\mathbf{a}}dx^{\nu}~~~h_{\mathbf{a}%
}^{\mu}h_{\mu}^{\mathbf{b}}=\delta_{\mathbf{a}}^{\mathbf{b}},~~~h_{\mathbf{a}%
}^{\mu}h_{\nu}^{\mathbf{a}}=\delta_{\nu}^{\mu} \label{8}%
\end{equation}
we immediately see that if
\begin{equation}
g_{\mu\nu}=\eta_{\mathbf{ab}}h_{\mu}^{\mathbf{a}}h_{\nu}^{\mathbf{a}%
},~~~~~g^{\mu\nu}=\eta^{\mathbf{ab}}h_{\mathbf{a}}^{\mu}h_{\mathbf{ab}}^{\mu}
\label{9}%
\end{equation}
it is
\begin{equation}
\boldsymbol{\eta}=\eta_{\mathbf{ab}}\mathfrak{g}^{\mathbf{a}}\otimes
\mathfrak{g}^{\mathbf{b}},~~~~~\mathtt{\eta}=\eta^{\mathbf{ab}}e_{\mathbf{a}%
}\otimes e_{\mathbf{b}}. \label{10}%
\end{equation}
and we have
\begin{equation}
D_{e_{\mathbf{a}}}e_{\mathbf{b}}=\omega_{\cdot\mathbf{ab}}^{\mathbf{c}%
\cdot\cdot}e_{\mathbf{c}},~~~~D_{e_{\mathbf{a}}}\mathfrak{g}^{\mathbf{c}%
}=-\omega_{\cdot\mathbf{ab}}^{\mathbf{c}\cdot\cdot}\mathfrak{g}^{\mathbf{b}},
\label{11a}%
\end{equation}
where the coefficients $\omega_{\cdot\mathbf{ab}}^{\mathbf{c}\cdot\cdot}$ of
the connection in the basis $\{e_{\mathbf{a}}\}$, $\{\mathfrak{g}^{\mathbf{b}%
}\}$ are called by physicists the spin connection. The reason for that name
will become clear in a while.

\textbf{8. }To proceed we recall that defining the objects\footnote{Take
notice that one can show that $\omega_{\cdot\mathbf{c\cdot}}^{\mathbf{a}%
\cdot\mathbf{b}}=-\omega_{\cdot\mathbf{c\cdot}}^{\mathbf{b}\cdot\mathbf{a}}$.}%
\begin{equation}
\omega_{\mathbf{c}}:=\frac{1}{2}\omega_{\cdot\mathbf{c\cdot}}^{\mathbf{a}%
\cdot\mathbf{b}}\mathfrak{g}^{\mathbf{a}}\mathfrak{g}^{\mathbf{b}}\in\sec%
{\textstyle\bigwedge\nolimits^{2}}
T^{\ast}M\hookrightarrow\sec\mathcal{C\ell}(M,\mathtt{\eta}) \label{11b}%
\end{equation}
we can easily show that for any $\mathcal{C\in}\sec\mathcal{C\ell
}(M,\mathtt{\eta})$ it is%
\begin{equation}
D_{e_{\mathbf{c}}}\mathcal{C}=\mathfrak{d}_{c}\mathcal{C+}\frac{1}{2}%
[\omega_{\mathbf{c}},\mathcal{C}], \label{11c}%
\end{equation}
where $\mathfrak{d}_{c}$ is the so-called Pfaff derivative of form fields. We
have%
\begin{equation}
\mathfrak{d}_{c}\mathcal{C}:=h_{\mathbf{c}}^{\mu}\frac{\partial}{\partial
x^{\mu}}\mathcal{C}. \label{11d}%
\end{equation}

\subsection{Dirac Operator Acting on $\mathcal{C\ell}(M,\mathtt{\eta})$}

\textbf{9. }The Dirac operator acting on the bundle of differential forms is
the invariant first order operator%
\begin{gather}
\boldsymbol{\partial}:=\mathfrak{g}^{\mathbf{a}}D_{e_{\mathbf{a}}}=\gamma
^{\mu}\mathrm{\partial}_{\mu}:\sec\mathcal{C\ell}(M,\mathtt{\eta}%
)\rightarrow\sec\mathcal{C\ell}(M,\mathtt{\eta}),\nonumber\\
\boldsymbol{\partial}\mathcal{C}=\boldsymbol{\partial}\wedge\mathcal{C}%
+\boldsymbol{\partial}\lrcorner\mathcal{C} \label{11}%
\end{gather}
where we can easily show that
\begin{equation}
\boldsymbol{\partial}\wedge\mathcal{C}=d\mathcal{C}\text{,~~~~~}%
\boldsymbol{\partial}\lrcorner\mathcal{C=-}\delta\mathcal{C,} \label{12}%
\end{equation}
where $d$ is the differential operator and $\delta$ is the Hodge
codifferential operator.

\subsection{Structure of the Clifford and spin-Clifford bundles}

\textbf{10.}The Clifford bundle (an algebra bundle) in a spin manifold (which
is the case of Minkowski spacetime and more generally of parallelizable
Lorentzian manifolds) is the vector bundle%
\begin{equation}
\mathcal{C\ell}(M,\mathtt{\eta})=P_{\mathrm{Spin}_{1,3}^{0}}\times
_{Ad}\mathbb{R}_{1.3} \label{13}%
\end{equation}
where $\mathbb{R}_{1.3}\simeq\mathbb{H(}2)$ is the so-called \emph{spacetime
algebra}. \ Also, $Ad:\mathrm{Spin}_{1,3}^{e}\rightarrow\mathrm{End(}%
\mathbb{R}_{1,3}\mathrm{)}$\textrm{ } is such that $Ad(u)a=uau^{-1}$. And
$\rho:\mathrm{SO}_{1,3}^{e}\rightarrow\mathrm{End(}\mathbb{R}_{1,3}\mathrm{)}$
is the natural action of $\mathrm{SO}_{1,3}^{e}$ on $\mathbb{R}_{1,3}$.

Take notice that the Dirac algebra is $\mathbb{R}_{4,1}\simeq\mathbb{C}(4)$
and that $\mathbb{R}_{4,1}\simeq\mathbb{C\otimes R}_{1.3}\simeq\mathbb{C}(4)$.

Now, the bundle of Dirac spinor fields\ is isomorphic to the
bundle\footnote{In Eq.(\ref{14}) $\mathrm{\ell}$ is the representation of
\textrm{Spin}$_{1,3}^{0}\simeq\mathrm{Sl}(2,\mathbb{C)}$\ on $I$ by left
multiplication.}%

\begin{equation}
S(M)=P_{\mathrm{Spin}_{1,3}^{0}}\times_{\mathrm{\ell}}I \label{14}%
\end{equation}
called the \emph{spin-Clifford bundle} where $I$ is the minimal left ideal
\begin{equation}
I=\mathbb{(C\otimes R}_{1.3})f \label{15}%
\end{equation}
generated by the idempotent%
\begin{equation}
f=\frac{1}{2}(1+\mathfrak{g}^{0})\frac{1}{2}(1+\mathrm{i}\mathfrak{g}%
^{2}\mathfrak{g}^{1}). \label{16}%
\end{equation}
Indeed, as well know, the bundle of Dirac spinor fields (as usually employed
in Physics texts books and papers) is the bundle\footnote{In EQ.(\ref{17})
$\mathbf{l}$ refers to the $\mathrm{D}^{1/2.0}\oplus\mathrm{D}^{1/2.0}$
representation of \textrm{Spin}$_{1,3}^{0}$ on the ideal $\mathbf{l}$.}
\begin{equation}
S_{D}(M)=P_{\mathrm{Spin}_{1,3}^{0}}\times_{\mathbf{L}}\mathbf{I} \label{17}%
\end{equation}
where $\mathbf{I}$ is the minimal left ideal
\begin{equation}
\mathbf{I}=\mathbb{C}_{4}\mathbf{f} \label{18}%
\end{equation}
generated by the idempotent
\begin{equation}
\mathbf{f}=\frac{1}{2}(1+\text{\ }\mathfrak{\underline{\gamma^{0}}})\frac
{1}{2}(1+\mathrm{i}\mathfrak{\underline{\gamma^{2}}\underline{\gamma^{1}}}).
\label{19}%
\end{equation}
with $\gamma^{\mathbf{a}}$ being the standard representation of the Dirac
gamma matrices. These objects are the matrix representations in $\mathbb{C}%
_{4}$ of the (orthonormal) 1-form fields $\mathfrak{g}^{\mathbf{a}}$

Now, recalling that the $S(M)$ is a module over $\mathcal{C\ell}%
(M,\mathtt{\eta})$ we observe that it is a nontrivial fact that once we fix a
spin-frame (i.e., an element of $P_{\mathrm{Spin}_{1,3}^{0}}$), say $\Xi_{0}$,
then any section of $\mathcal{S}(M)$ can be written any $\Psi\in\sec S(M)$ can
be written as
\begin{equation}
\Psi_{\Xi_{0}}=\psi_{\Xi_{0}}\frac{1}{2}(1+\mathfrak{g}^{0})\frac{1}%
{2}(1+\mathrm{i}\mathfrak{g}^{2}\mathfrak{g}^{2}) \label{20}%
\end{equation}
where $\psi\in\sec\mathcal{C\ell}^{0}(M,\mathtt{\eta})$, with $\mathcal{C\ell
}^{0}(M,\mathtt{\eta})$ the even subalgebra of $\mathcal{C\ell}(M,\mathtt{\eta
})$. Then, a general $\psi$ can be \ (conveniently written , for what follows)
as
\begin{equation}
\psi_{\Xi_{0}}=-\mathcal{S}+\mathcal{F-\gamma}^{5}\mathcal{P}, \label{21}%
\end{equation}
with%
\begin{align}
\mathcal{S}  &  \in\sec%
{\textstyle\bigwedge\nolimits^{0}}
T^{\ast}M\hookrightarrow\sec\mathcal{C\ell}^{0}(M,\mathtt{\eta}),\nonumber\\
\mathcal{F}  &  \in\sec%
{\textstyle\bigwedge\nolimits^{2}}
T^{\ast}M\hookrightarrow\sec\mathcal{C\ell}^{0}(M,\mathtt{\eta}),\nonumber\\
\mathcal{\gamma}^{5}\mathcal{P}  &  \mathcal{\in}\sec%
{\textstyle\bigwedge\nolimits^{4}}
T^{\ast}M\hookrightarrow\sec\mathcal{C\ell}^{0}(M,\mathtt{\eta}) \label{22}%
\end{align}
where we have use the notable fact that for any $\mathcal{C}\in\sec
\mathcal{C\ell}(M,\mathtt{\eta})$ we can write the Hodge star operator as%
\begin{equation}
\star\mathcal{C}=\mathcal{\tilde{C}\gamma}^{5} \label{23}%
\end{equation}
with $\mathcal{\gamma}^{5}=\mathfrak{g}^{0}\wedge\mathfrak{g}^{1}%
\wedge\mathfrak{g}^{1}\wedge\mathfrak{g}^{3}=\mathfrak{g}^{0}\mathfrak{g}%
^{1}\mathfrak{g}^{1}\mathfrak{g}^{3}\mathcal{\in}\sec%
{\textstyle\bigwedge\nolimits^{4}}
T^{\ast}M\hookrightarrow\sec\mathcal{C\ell}^{0}(M,\mathtt{\eta})$ is the
volume element.

\begin{remark}
The object $\psi_{\Xi_{0}}$ is said to be a \emph{representative} in the
Clifford bundle and in the spin frame $\Xi_{0}$\ of a Dirac-Hestenes spinor.
Thus a Dirac-Hestenes spinor field is a certain equivalence class of even
sections of the Clifford bundle. If you want to see details, please consult
\cite{rc2007}.
\end{remark}

\begin{remark}
To save notation in what follows we will simply write $\Psi$ and $\psi$ for
$\Psi_{\Xi_{0}}$ and $\psi_{\Xi_{0}}$.
\end{remark}

\subsection{The spin-Dirac Operator}

\textbf{11.}The spin-Dirac operator is the first order differential operator
\begin{equation}
\partial^{s}:=\mathfrak{g}^{\mathbf{a}}D_{\mathbf{e}_{a}}^{s}=\mathfrak{\gamma
}^{\mu}\mathrm{\partial}_{\mu}:\sec S(M)\rightarrow\sec S(M), \label{24}%
\end{equation}
where $D^{s}$ is the spinor covariant derivative (spin connection) such that
for each $\Psi\in\sec S(M)$ it is
\begin{equation}
D_{\mathbf{e}_{a}}^{s}\Psi:=e_{\mathbf{a}}(\Psi)+\frac{1}{2}\omega
_{\mathbf{a}}\Psi\label{25}%
\end{equation}

Note that the cobasis $\{\mathfrak{\gamma}^{\mu}\}$ is orthonormal and for
that basis $\omega_{\mu}=0$ and then in this basis
\begin{equation}
\partial^{s}=\mathfrak{\gamma}^{\mu}D_{\mathrm{\partial}_{\mu}}^{s}%
=\mathfrak{\gamma}^{\mu}\mathrm{\partial}_{\mu}. \label{25a}%
\end{equation}

Now $D_{\mathbf{e}_{a}}^{(s)}$, the representative\ of $D_{\mathbf{e}_{a}}%
^{s}$ in the Clifford bundle acts on representatives of Dirac-Hestenes spinor
fields $\psi\in\sec\mathcal{C\ell}^{0}(M,\mathtt{\eta})$ as%
\begin{equation}
D_{\mathbf{e}_{a}}^{s}\psi:=\mathfrak{d}_{\mathbf{a}}(\psi)+\frac{1}{2}%
\omega_{\mathbf{a}}\psi\label{26}%
\end{equation}
Of course, we also have a representative $\partial^{(s)}$ of the spin-Dirac
operator acting on the representative of Dirac-Hestenes spinor fields $\psi
\in\sec\mathcal{C\ell}^{0}(M,\mathtt{\eta})$. It is:%
\[
\partial^{(s)}\psi=\mathfrak{g}^{\mathbf{a}}D_{\mathbf{e}_{a}}^{(s)}\psi
\]

\begin{remark}
So, using the cobasis $\{\mathfrak{\gamma}^{\mu}\}$ the expressions for the
Dirac operator $\partial$ and the spin-Dirac operator are the same, namely
$\partial=\mathfrak{\gamma}^{\mu}\mathrm{\partial}_{\mu}$ and $\partial
^{s}=\mathfrak{\gamma}^{\mu}\mathrm{\partial}_{\mu}$. So, in what follows
since we are only going to use the orthonormal cobasis $\{\mathfrak{\gamma
}^{\mu}\}$ we will denote both as $\boldsymbol{\partial}$ when this does not
cause any misunderstanding.
\end{remark}

\subsection{The Dirac-Hestenes Equation}

\textbf{12. }The Dirac equation is represented in the Clifford bundle by a
representative of a Dirac-Hestenes spinor field once a spin frame is fixed. We
have, as well known \cite{hestenes, rc2007}
\begin{equation}
\partial\psi\gamma^{21}-m\psi\gamma^{0}=0. \label{27}%
\end{equation}
If we multiply this equation on the right by the idempotent $f$ we immediately
get the following equation for $\Psi,$%
\begin{equation}
\mathrm{i}\gamma^{\mu}\mathrm{\partial}_{\mu}\Psi-m\Psi=0 \label{28}%
\end{equation}
This equation in the bundle $S_{D}(M)$ is for $\mathbf{\Psi}\in\sec S_{D}(M)$
\begin{equation}
\mathrm{i}\text{\ }\mathfrak{\underline{\gamma}}^{\mu}\mathrm{\partial}_{\mu
}\mathbf{\Psi}-m\mathbf{\Psi}=0 \label{29}%
\end{equation}
where since $S_{D}(M)$ is trivial we can take (once a spin frame is chosen)
\begin{equation}
\mathbf{\Psi}:\mathbb{R}^{4}\rightarrow\mathbb{C}^{4}. \label{30}%
\end{equation}

\section{Generalized Maxwell Equation in the Clifford Bundle}

The generalized Maxwell equation (GME) for $F=\frac{1}{2}F_{\mu\nu}\gamma
^{\mu}\wedge\gamma^{v}$ $\in\sec%
{\textstyle\bigwedge\nolimits^{2}}
T^{\ast}M\hookrightarrow\sec\mathcal{C\ell}(M,\mathtt{\eta})$ generated by an
electric current $J_{e}\in\sec%
{\textstyle\bigwedge\nolimits^{1}}
T^{\ast}M\hookrightarrow\sec\mathcal{C\ell}(M,\mathtt{\eta})$ and an magnetic
current $J_{m}\in\sec%
{\textstyle\bigwedge\nolimits^{1}}
T^{\ast}M\hookrightarrow\sec\mathcal{C\ell}(M,\mathtt{\eta})$ can be written
in the Clifford bundle formalism as%
\begin{equation}
\partial F=J_{e}+\gamma^{5}J_{m}. \label{36}%
\end{equation}
Indeed, taking into account that $\star J_{m}=J_{m}\gamma^{5}=-\gamma^{5}%
J_{m}$ and that $\partial=d-\delta$ Eq.(\ref{36}) is equivalent to the
following pair of equations%
\begin{equation}
\delta F=-J_{e},~~~~~dF=\gamma^{5}J_{m}=-\star J_{m} \label{37}%
\end{equation}

Note that the equation $dF=\gamma^{5}J_{m}=-\star J_{m}$ can be written as%
\begin{equation}
\delta\star F=J_{m}\text{.} \label{38}%
\end{equation}
Obviously, a field $F\in\sec%
{\textstyle\bigwedge\nolimits^{2}}
T^{\ast}M\hookrightarrow\sec\mathcal{C\ell}(M,\mathtt{\eta})$ satisfying
Eq.(\ref{36}) cannot be derived form a potential $A\in\sec%
{\textstyle\bigwedge\nolimits^{1}}
T^{\ast}M\hookrightarrow\sec\mathcal{C\ell}(M,\mathtt{\eta})$. However, $F$
can be derived from the superpotential\footnote{Known as Cabibo-Ferrari
potential. See more details in \cite{mrrr}.
\par
{}}%
\begin{align}
\mathcal{A}  &  :=A+\gamma^{5}B,\label{39}\\
A,B  &  \in\sec%
{\textstyle\bigwedge\nolimits^{1}}
T^{\ast}M\hookrightarrow\sec\mathcal{C\ell}(M,\mathtt{\eta})
\end{align}
if we impose the Lorenz gauge to $A$ and $B$, i.e.
\begin{equation}
\delta A=-\partial\lrcorner A=0,~~~\delta B=-\partial\lrcorner B=0. \label{40}%
\end{equation}

Indeed, in this case%
\begin{equation}
F=\partial\mathcal{A}=\partial\wedge A+\partial\lrcorner A+\partial
\wedge(\gamma^{5}B)+\partial\lrcorner(\gamma^{5}B). \label{41}%
\end{equation}

Now,%
\begin{gather}
\partial\wedge(\gamma^{5}B)=\gamma^{\mu}\wedge\mathrm{\partial}_{\mu}%
(\gamma^{5}B)=\gamma^{\mu}\wedge(\gamma^{5}\mathrm{\partial}_{\mu
}B)\nonumber\\
=\frac{1}{2}(\gamma^{\mu}\gamma^{5}\mathrm{\partial}_{\mu}B-\gamma
^{5}\mathrm{\partial}_{\mu}B\gamma^{\mu})\nonumber\\
=-\gamma^{5}\frac{1}{2}(\gamma^{\mu}\mathrm{\partial}_{\mu}B+\mathrm{\partial
}_{\mu}B\gamma^{\mu})=-\gamma^{5}\partial\lrcorner B=-\star\partial\lrcorner
B=\star\delta B. \label{42}%
\end{gather}
On the other hand,%
\begin{align}
\partial\lrcorner(\gamma^{5}B)  &  =\gamma^{\mu}\lrcorner\mathrm{\partial
}_{\mu}(\gamma^{5}B)=\gamma^{\mu}\lrcorner(\gamma^{5}\mathrm{\partial}_{\mu
}B)\nonumber\\
&  =\frac{1}{2}(\gamma^{\mu}\gamma^{5}\mathrm{\partial}_{\mu}B+\gamma
^{5}\mathrm{\partial}_{\mu}B\gamma^{\mu})\nonumber\\
&  =-\gamma^{5}\frac{1}{2}(\gamma^{\mu}\mathrm{\partial}_{\mu}%
B-\mathrm{\partial}_{\mu}B\gamma^{\mu})=-\gamma^{5}\partial\wedge
B=\star\partial\wedge B=\star dB. \label{43}%
\end{align}

So, with conditions given in (\ref{40}) we have%
\begin{equation}
F=\partial\mathcal{A}=dA+\star dB \label{44}%
\end{equation}
Then,%
\begin{equation}
\partial F=(d-\delta)(dA+\star dB)=d\star dB-\delta dA \label{45}%
\end{equation}
and we must have
\begin{equation}
J_{e}=-\delta dA,~~~~~\star J_{m}=d\star dB. \label{46}%
\end{equation}

Moreover, since $A$ and $B$ satisfy the Lorenz condition we have that they
obey nonhomogeneous wave equations. Indeed,%
\begin{align}
-\delta dA-\delta dA  &  =(d-\delta)^{2}A=\partial^{2}A=\partial\wedge\partial
A+\partial\cdot\partial A\label{47}\\
\partial\cdot\partial A  &  =J_{e}%
\end{align}
since $\partial\wedge\partial A=0$. Also, it is
\begin{equation}
\partial\cdot\partial B=J_{e}. \label{49}%
\end{equation}

\section{The Neutrino Equation in Maxwell Like Form}

A massless neutrino is represented by a Weyl spinor field. In the Clifford
bundle formalism a representative of a Weyl spinor field is
\begin{equation}
\phi=\frac{1}{2}\psi(1+\gamma^{5}) \label{50}%
\end{equation}
where $\psi$ is a representative of a Dirac-Hestenes spinor field \ in
$\mathcal{C\ell}(M,\mathtt{\eta})$. Now, if
\begin{equation}
\boldsymbol{\partial}\psi=0, \label{50a}%
\end{equation}
(massless Dirac-Hestenes Equation) $\phi$ satisfy the Weyl equation%
\begin{equation}
\boldsymbol{\partial}\phi=0, \label{51}%
\end{equation}
So, it is enough to show how to write the Eq.(\ref{50a}) as a GME. Indeed,
recalling Eq.(\ref{21}) we immediately see that we can write
\begin{equation}
\boldsymbol{\partial}\psi=-\boldsymbol{\partial}\mathcal{S}%
+\boldsymbol{\partial}\mathcal{F-\boldsymbol{\partial}\gamma}^{5}\mathcal{P}=0
\label{52}%
\end{equation}
and thus we have
\begin{equation}
\boldsymbol{\partial}\mathcal{F}=\boldsymbol{\partial}%
\mathcal{S+\mathcal{\gamma}}^{5}\mathcal{\boldsymbol{\partial}P} \label{54}%
\end{equation}
and calling
\begin{align}
\mathcal{J}_{e}  &  :=\boldsymbol{\partial}\mathcal{S\in}\sec%
{\textstyle\bigwedge\nolimits^{1}}
T^{\ast}M\hookrightarrow\sec\mathcal{C\ell}(M,\mathtt{\eta}),\nonumber\\
\mathcal{J}_{m}  &  :=\mathcal{\boldsymbol{\partial}P\in}\sec%
{\textstyle\bigwedge\nolimits^{1}}
T^{\ast}M\hookrightarrow\sec\mathcal{C\ell}(M,\mathtt{\eta}), \label{55}%
\end{align}
we see that Eq.(\ref{54}) can be written as%
\begin{equation}
\boldsymbol{\partial}\mathcal{F}=\mathcal{J}_{e}\mathcal{+\mathcal{\gamma}%
}^{5}\mathcal{J}_{m} \label{56}%
\end{equation}
which is formally identical to Eq.(\ref{36}), the GME generated by electric
and magnetic currents.

\section{Bosonization of Fermionic Fields}

Does the steps we use to go from Eq.(\ref{50a}) to Eq.(\ref{56}) means that
we\ \textquotedblleft bosonized\textquotedblright\ a spinor field?

The answer is \emph{yes }in the following sense\emph{.} First recall that the
spin-Dirac operator and the Dirac operator (which act on very different
bundles) have the same form $\gamma^{\mu}\mathrm{\partial}_{\mu}$ only because
we used coordinates for $M$ in Einstein-Lorentz-Poincar\'{e} gauge and
Eqs.(\ref{50a}) and (\ref{51}) are only representatives of equivalent
differential equations for legitimate spinor fields. This means that for%
\begin{equation}
\Psi=\psi f\in\sec S(M) \label{57}%
\end{equation}
we have that%
\begin{equation}
\boldsymbol{\partial}\Psi=0 \label{58}%
\end{equation}
which using the decomposition given by Eq.(\ref{21}) we get
\begin{equation}
\boldsymbol{\partial}\mathcal{F}f=\mathcal{J}_{e}f\mathcal{+\mathcal{\gamma}%
}^{5}\mathcal{J}_{m}f \label{59}%
\end{equation}
where $\mathcal{F}f,\mathcal{J}_{e}f,\mathcal{\mathcal{\gamma}}^{5}%
\mathcal{J}_{m}f\in\sec S(M)$, i.e., they are spinor fields.

However, we already remarked that Eq.(\ref{59}) is the expression of the Weyl
equation once a spin frame is fixed. For different spin frames the object that
represent the intrinsic object called a spinor field in the Clifford bundle is
represented by different sections of the Clifford bundle. Indeed, if
$\psi=\psi_{\Xi_{0}}$ is the representative of a spinor field in the spin
frame $\Xi_{0}$ and $\psi_{\Xi}$ is the representative of the same spinor
field in the spin frame $\Xi$, then we have%
\begin{equation}
\psi_{\Xi_{0}}u_{0}=\psi_{\Xi}u \label{60}%
\end{equation}
i.e., a spinor field, (a \emph{fermion} field) which is a section of a spinor
bundle $S(M)$ may be represented by an equivalence class of \ nonhomogeneous
even sections of the Clifford bundle\footnote{Details may be found in
\cite{mr2004,r2004,rc2007}.} (room of \emph{boson }fields). For the particular
case of the Weyl spinor field satisfying Weyl equation we can say that the
Weyl spinor field is equivalent to an equivalent class of scalar, plus 2-forms
and \ plus 4-forms fields, namely, $\mathcal{S},\mathcal{F},\gamma
^{5}\mathcal{P}$, coupled through Eq.(\ref{54}).

\section{The Electron Equation in Maxwell Form}

Introduce a Hertz potential field \cite{rc2007,stratton} $\Pi\in\sec
\bigwedge\nolimits^{2}T^{\ast}M\hookrightarrow\sec\mathcal{C\ell
}(M,\mathtt{\eta})$ satisfying the following equation
\begin{equation}
\boldsymbol{\partial}\Pi=(\boldsymbol{\partial}\mathfrak{G}+m\mathfrak{P}%
\gamma_{3}+m\langle\Pi\gamma_{012}\rangle_{1})+\gamma_{5}(\boldsymbol{\partial
}\mathfrak{P}+m\mathfrak{G}\gamma_{3}-\gamma_{5}\langle m\Pi\gamma
_{012}\rangle_{3}) \label{mde7.1}%
\end{equation}
where $\mathfrak{G,P}\in\sec\bigwedge\nolimits^{0}T^{\ast}M\hookrightarrow
\sec\mathcal{C\ell}(M,\mathtt{\eta})$, and $m$ is a constant. Under these
conditions, the electromagnetic and Stratton potentials \cite{rc2007} are
\begin{equation}
A=\boldsymbol{\partial}\mathfrak{G}+m\mathfrak{P}\gamma_{3}+m\langle\Pi
\gamma_{012}\rangle_{1,} \label{mde7.2}%
\end{equation}%
\begin{equation}
\gamma_{5}S=\gamma_{5}(\boldsymbol{\partial}\mathfrak{P}+m\mathfrak{G}%
\gamma_{3}-\gamma_{5}\langle m\Pi\gamma_{012}\rangle_{3}), \label{mde7.3}%
\end{equation}
and must thus satisfy the following subsidiary conditions,
\begin{equation}
\Diamond(\boldsymbol{\partial}\mathfrak{G}+m\mathfrak{P}\gamma_{3}+m\langle
\Pi\gamma_{012}\rangle_{1})=J_{e}, \label{mde7.4}%
\end{equation}%
\begin{equation}
\Diamond(\gamma_{5}(\boldsymbol{\partial}\mathfrak{P}+m\mathfrak{G}\gamma
_{3}-\gamma_{5}\langle m\Pi\gamma_{012}\rangle_{3}))=0, \label{mde7.5}%
\end{equation}%
\begin{equation}
\Diamond\mathfrak{G}+m\boldsymbol{\partial}\cdot\langle\Pi\gamma_{012}%
\rangle_{1}=0, \label{mde7.6}%
\end{equation}%
\begin{equation}
\Diamond\mathfrak{P}-m\boldsymbol{\partial}\cdot(\gamma_{5}\langle\Pi
\gamma_{012}\rangle_{3})=0, \label{mde7.7}%
\end{equation}
where $\Diamond=-(d\delta+\delta d)=\boldsymbol{\partial}^{2}%
=\boldsymbol{\partial}\wedge\boldsymbol{\partial+\partial}\lrcorner
\boldsymbol{\partial}$.

Now, in the Clifford bundle formalism, the following sum is a legitimate
operation
\begin{equation}
\psi=-\mathfrak{G}+\Pi+\gamma_{5}\mathfrak{P} \label{mde7.8}%
\end{equation}
and according to previous results Eq.(\ref{mde7.8}) defines $\psi$ as a
representative of \ some Dirac-Hestenes spinor field in a given spin frame.
Now, we can verify that $\psi$ satisfies the equation
\begin{equation}
\boldsymbol{\partial}\psi\gamma_{21}-m\psi\gamma_{0}=0 \label{mde7.9}%
\end{equation}
which is as we already know a \emph{representative} of the standard Dirac
equation (for a free electron) in the Clifford bundle (Eq.(\ref{27})). Once
again we can say that we bosonized the electron field.

\section{The Fermionization of Maxwell Field}

As we already said, in the Clifford bundle the (generalized) Maxwell equation
satisfied by $F\in\sec%
{\textstyle\bigwedge\nolimits^{2}}
T^{\ast}M\hookrightarrow$ $\sec\mathcal{C\ell}(M,\mathtt{\eta})$ is
\begin{equation}
\partial F=J_{e}+\gamma^{5}J_{m}. \label{80}%
\end{equation}
with electric current $J_{e}\in\sec%
{\textstyle\bigwedge\nolimits^{1}}
T^{\ast}M\hookrightarrow\sec\mathcal{C\ell}(M,\mathtt{\eta})$ and an magnetic
current $J_{m}\in\sec%
{\textstyle\bigwedge\nolimits^{1}}
T^{\ast}M\hookrightarrow\sec\mathcal{C\ell}(M,\mathtt{\eta})$. Now, recall
that the spin-Clifford bundle is module over the Clifford
bundle\cite{cru,rc2007}. Then, choosing a spin frame, say $\Xi_{0}$ and an
idempotent field $f\in\sec S(M)$ and multiplying Eq.(\ref{80}) on the right by
$f$ we get for $\Psi_{\Xi_{0}}=Ff\in\sec S(M)$ the equation
\begin{equation}
\partial\Psi_{\Xi_{0}}=\mathcal{J}_{e\Xi_{0}}+\gamma^{5}\mathcal{J}_{m\Xi_{0}%
}, \label{81}%
\end{equation}
where $\mathcal{J}_{e\Xi_{0}}=J_{e}f,\mathcal{J}_{m\Xi_{0}}=J_{m}f\in\sec
S(M)$. In this way we can say that the Maxwell field satisfying Maxwell
equation is an equivalence class of fermion fields $[\Psi_{\Xi_{0}}]$ and we
can say that we fermionize a bosonic field!.

\begin{remark}
Of, course, there several different spinor like representations of the Maxwell
field, since there are many non equivalent idempotent fields in $S(M)$. By
choosing appropriately these idempotents we can reproduce all Dirac like
representations that appeared in the literature. Details in
\emph{\cite{rc1990}}
\end{remark}

\section{Conclusions}

In this brief note we showed how using the Clifford and spin-Clifford bundles
formalism we can give a rigorous mathematical meaning to the meaning of the
sentences: (i) bosonization of fermionic fields and (ii) fermionization of
bosonic fields. Each object in one class is represented by a well defined
equivalence class of objects in the other class. It is also important to take
in mind that the mathematical equivalences proved above do not imply, of
course, any change in the statistics \ satisfied by each class (fermion or
boson) of fields. We finally recall that the above formalism can be trivially
generalized for fields living in a general Lorentzian spacetime structure, but
in this case care must be taken in distinguishing explicitly the Clifford and
spin-Clifford Dirac operators.

\begin{acknowledgement}
The author thanks Professor L.C.L. Botelho for his suggestion to write this
paper and his very important comments.
\end{acknowledgement}


\begin{thebibliography}{9}                                                                                                %
\bibitem {cru}{\footnotesize Crumeyrolle, A., \emph{Orthogonal and Symplectic
Clifford Algebras. Spinor Structures}. Kluwer Acad. Publ.,,Dordrecht, 1990.}

\bibitem {hestenes}{\footnotesize Hestenes, D., \emph{Space-Time Algebra}
(second revised edition), Birkh\"{a}user, Basel, 2015.}

\bibitem {mrrr}{\footnotesize Maia, A. Jr.,Recami E. , Rodrigues, W. A. Jr.,
and Rosa, M. A. F., Magnetic Monopoles without String in the
K\"{a}hler-Clifford Algebra Bundle: A Geometrical Interpretation, \emph{J.
Math.Phys. }\textbf{31}, 502-505 (1990).}

\bibitem {rc1990}{\footnotesize Rodrigues, W. A. Jr. and Capelas de Oliveira,
E., Dirac and Maxwell Equations in the Clifford and Spin-Clifford bundles,
\emph{Int. J. Theor. Phys}. \textbf{29}, 397-412 (1990).}

\bibitem {mr2004}{\footnotesize Mosna, R. A., and Rodrigues, W. A. Jr., The
Bundles of Algebraic and Dirac-Hestenes Spinor Fields, \emph{J. Math. Phys}.
\textbf{45}, 2945-2966 (2004).}

\bibitem {r2004}{\footnotesize Rodrigues, W. A. Jr., Algebraic and
Dirac-Hestenes Spinors and Spinor Fields, \emph{J. Math. Phys}. \textbf{45},
2908-2994 (2004)}.

\bibitem {rc2007}{\footnotesize Rodrigues, W. A. Jr. and Capelas de Oliveira,
E., \emph{The Many Faces of Maxwell Dirac and Einstein Equations. A Clifford
Bundle Approach}, Lecture Notes in Physics \textbf{922} (second edition
revised and enlarged), Springer, Heidelberg, 2016 \ (first published as
Lecture Notes in Physics \textbf{722}, 2007).}

\bibitem {stratton}{\footnotesize Stratton, J. A., \emph{Electromagnetic
Theory}, McGraw-Hill Book Co., New York, 1941}
\end{thebibliography}
\end{document}